\documentclass[twocolumn,amsmath,amssymb]{revtex4}

\usepackage{graphicx}
\usepackage{dcolumn}
\usepackage{bm}

\begin{document}


\title{L\'evy flights of photons in hot atomic vapours}

\author{N. Mercadier$^1$}
\author{W. Guerin$^1$}%
\author{M. Chevrollier$^2$}
\author{R. Kaiser$^1$}
 \email{Robin.Kaiser@inln.cnrs.fr}
\affiliation{%
$^1$Institut Non Lin\'eaire de Nice, CNRS and Universit\'e de Nice
Sophia-Antipolis, 1361 route des Lucioles,
06560 Valbonne, France,\\
$^2$Universidade Federal da Paraiba, Cx. Postal 508, 58051-900
Joao Pessoa-PB, Brazil
}%

\begin{abstract}
Properties of random and fluctuating systems are often studied
through the use of Gaussian distributions. However, in a number of
situations, rare events have drastic consequences, which can not
be explained by Gaussian statistics. Considerable efforts have
thus been devoted to the study of non Gaussian fluctuations such
as L\'evy statistics, generalizing the standard description of
random walks. Unfortunately only macroscopic signatures, obtained
by averaging over many random steps, are usually observed in
physical systems. We present experimental results investigating
the elementary process of anomalous diffusion of photons in hot
atomic vapours. We measure the step size distribution of the
random walk and show that it follows a power law characteristic of
L\'evy flights.
\end{abstract}

\date{\today}

\maketitle

Random walk of particles in a disordered or fluctuating medium is
often well described by a diffusion equation, characterized by a
linear increase in time of the mean square displacement of the
particles: $<r^2>=Dt$, with $D$ the diffusion coefficient. One
assumption for the diffusion equation to hold is that the size $x$
of each step of the random walk is given by a distribution $P(x)$
with a finite second moment $<x^2>$, allowing to apply the central
limit theorem. When the step size distribution $P(x)$ follows an
asymptotic power law $P(x) \sim 1/x^\alpha$, the moments of the
distributions can however become infinite.  It has been long
established that for $\alpha<3$, the average square displacement
is governed by rare but large steps \cite{Levy}. Such a class of
random walk is called L\'evy flights, corresponding to a
superdiffusive behaviour where $<r^2>=Dt^\gamma$, with $\gamma>1$.
The broad range of applications of L\'evy flights includes
biology, economics, finance, catastrophe management and resonance
fluorescence in astrophysical systems and atomic vapours
\cite{Bouchaud,Shlesinger,Metzler,Bouchaud2}. Large (non Gaussian)
fluctuations also play a fundamental role in many physical
situations, in particular around phase transitions, having
triggered considerable efforts to understand universal features of
such phenomena \cite{Botet,Goldenfeld}.

Anomalous transport of photons has been reported recently in
engineered optical material \cite{Wiersma}. Superdiffusive
behavior of light has also been known in the context of radiation
trapping in hot atomic vapors. Because this phenomenon occurs in
many different systems, ranging from stars \cite{Springmann} to
dense atomic vapours \cite{Molisch} such as gas lasers, discharges
and hot plasmas, this field has been subject to intense studies
for many decades, including seminal work by Holstein
\cite{Holstein}. It has been realized very early \cite{Kenty} that
frequency redistribution has a profound impact on the multiple
scattering features of light. Whereas elastic scattering, which
occurs in laser-cooled dilute atomic vapours
\cite{Fioretti,Labeyrie2003,Labeyrie2005}, leads to normal
diffusion with well defined scattering mean free path and
diffusion coefficient, inelastic scattering as in hot vapours can
lead to situations where the central limit theorem no longer
applies and photon trajectories are expected to be L\'evy flights.
Unfortunately, in most systems where multiple scattering of light
occurs, it is difficult to have direct experimental access to the
single step size distribution at the origin of the random walk and
anomalous diffusion is usually inferred from macroscopic
observations \cite{Molisch, Wiersma}.

In this Article we present experimental results investigating the
microscopic ingredient leading to a regime of superdiffusion in
the multiple scattering of light in hot vapours of rubidium atoms.
We have used a specific geometrical arrangement to isolate a
single step in the multiple scattering sequence. We measure the
single step size distribution $P(x)$, which follows a power law
$P(x)\propto\,1/x^\alpha$, with $\alpha<3$. Therefore the photon
trajectories are L\'evy flights, with an infinite variance of
$P(x)$.

The random walk of light in atomic vapours is usually
characterized by the various moments of the step size distribution
$P(x)$. One can define a mean free path by $\ell=\langle
x\rangle=\int_0^\infty\, x P(x)dx$ and a diffusion coefficient
from the variance $\sigma_x^2$ of the distribution. For photons at
frequency $\omega$, this step size distribution $P(x, \omega)$ is
deduced from $P(x, \omega)=-\frac{\partial T(x, \omega)}{\partial
x}$, where $T$ is the forward transmission given by the
Beer-Lambert's law $T(x, \omega) =  e^{- x / \ell(\omega)}$, with
the frequency dependant mean free path $\ell(\omega)$. This
results in an exponential distribution
\begin{equation}\label{eq:Pxw}
P(x,\omega) = \frac{1}{\ell(\omega)} e^{- x/\ell(\omega)} \; .
\end{equation}

While these considerations give a good description for atoms close
to zero temperature \cite{Fioretti,Labeyrie2003,Labeyrie2005},
most samples in and outside laboratories present a Doppler
broadening $\Delta\omega_\mathrm{D}$ much larger than the natural
linewidth $\Gamma$ of the optical transition. The normalized
spectrum $\Theta(\omega)$ of the light in the multiple scattering
regime then influences the properties of its random walk
\cite{Kenty}. For instance photons with a frequency in the wings of the absorption
line will travel over a much larger distance than photons at the center of the absorption line.
The step size distribution $P(x)$ is obtained by
a frequency average of $P(x,\omega)$ [Eq. (\ref{eq:Pxw})] weighted
by the spectrum $\Theta(\omega)$:
\begin{equation}
\label{eq:P} P(x)= \int_0^{+\infty} \Theta(\omega) \times
\frac{1}{\ell(\omega)} e^{- x/\ell(\omega)} d\omega \; .
\end{equation}
The subsequent single step size distribution of light in an atomic
vapour can be numerically computed given the knowledge of the
emission spectrum $\Theta(\omega)$ and of the absorption spectrum
$1/\ell(\omega)$ (see the Methods section). An analytical
expression can be obtained in some limiting cases. For instance,
assuming that the emission and absorption spectra are purely
Gaussian (as given by pure Doppler broadening and neglecting the
atomic natural Lorentzian absorption line), then the single step
size distribution asymptotically follows \cite{Holstein,Pereira}:
\begin{equation}\label{eq.P2}
P(x) \sim \frac{1}{x^2 \sqrt{\ln(x/\ell_0)}} \; ,
\end{equation}
where $\ell_0^{-1}=n_\mathrm{at}\sigma_0$ with $n_\mathrm{at}$ the
atomic density and $\sigma_0$ the scattering cross section at the
atomic resonance. This heavy tail distribution leads to L\'evy
flights of light. As explained above, the underlying mechanism of
this power law is a frequency average of the Beer-Lambert
transmission, as different frequency components are not scattered
after the same distance.

To measure this step size distribution, we have used a specific
multicell arrangement as shown in Fig. 1. We image on a cooled
charge coupled camera (CCD) the fluorescence of a natural isotopic
mixture of rubidium atoms in a long cylindrical observation cell,
illuminated along its axis. We thus measure the probability of a
photon to be scattered after a distance $x$ along the cell axis,
\textit{i.e.} the step size distribution $P(x)$.

As a reference, we first measure $P(x)$ in the case when the
observation cell is illuminated by a monochromatic incident laser,
locked to the $F=3 \rightarrow F'=4$ transition of the D2 line of
rubidium 85. From the corresponding image [Fig. 2(a)] we extract
an exponential step size distribution as expected from
Beer-Lambert's law [Eq. (\ref{eq:Pxw}) \& Fig. 3]. This
preliminary measurement allows the calibration of the mean free
path for resonant light and thus the atomic density, which can be
varied by adjusting the temperature of the observation cell from
$20^{\circ}$C to about $47^\circ$C. The atomic density thus varies
from $9 \times 10^{15}$ to $2 \times 10^{17}$~m$^{-3}$, and the
mean free path changes accordingly from 50~mm to 5~mm.

\begin{figure}[t]
\begin{center}
\includegraphics{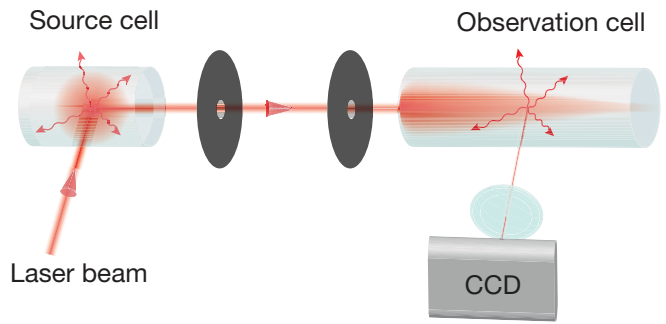}
\caption{{\bf The experimental setup.} A laser beam is incident on
a so-called source cell filled with rubidium vapour. Scattered
light propagating at orthogonal direction is selected with two
diaphragms and illuminates a second, observation cell. The light
scattered in this second cell is imaged on a cooled CCD camera.
This fluorescence signal is proportional to the step size
distribution function.}
\end{center}
\end{figure}

\begin{figure}[b]
\begin{center}
\includegraphics{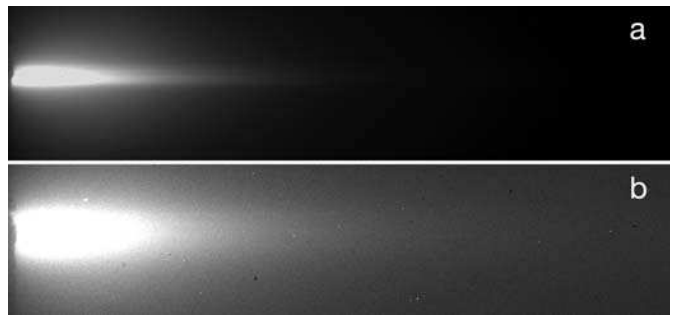}
\caption{{\bf Fluorescence images} Data are obtained after an
exposure time of 30 minutes and dark frame subtraction, for (a) an
incident laser beam at the atomic resonance frequency and (b)
incident light provided by a first scattering cell. The
temperature of the observation cell is $41^\circ$C. The step size
distribution is extracted from the intensity along the axis.}
\end{center}
\end{figure}

To measure L\'evy flights of light we use a double cell
configuration [Fig. 1]. Our 11~mW, 2~mm-waist laser beam is
incident on a first, small cylindrical cell of rubidium of optical
thickness $0.2$, where photons undergo at most one scattering
event with a well defined position. A 2~mm-diameter pencil of
diffused light propagating in a direction orthogonal to the
initial laser beam is then selected by two 12~cm-spaced
diaphragms.  This scheme produces photons with a frequency
spectrum $\Theta(\omega)$ given by the Doppler broadening. If we
neglect the finite but small width of the atomic transition, the
absorption spectrum is also purely Doppler, and Eq. (\ref{eq.P2})
holds. The scattered light then goes through the 7~cm-long
observation cell, with an angle of about $10^\circ$ from the cell
axis to avoid stray reflections at the center of the image. Raw
images of the fluorescence signal are obtained after a 30 minutes
exposure. Reproducible noise is then eliminated by subtracting a
dark frame. The resulting image is shown in Fig. 2(b). We extract
the corresponding step size distribution $P(x)$ (shown on a
log-log scale in Fig. 3) by taking longitudinal slices along the
incident direction of propagation. This signal is integrated over
30 lines of the CCD matrix (corresponding to 1.6~mm in the cell),
then smoothened over 30 pixels along the $x$ direction to increase
the signal to noise ratio. In order to obtain the correct $P(x)$
distribution, we need to correct the effect of multiple scattering
on the signal. We thus subtract the intensity measured along a
slice slightly off the line of sight of the diaphragms, which is
due only to multiple scattering, from the intensity measured on
the central slice (see the Methods section). We clearly identify a
power law, which can be fitted by:
\begin{equation}\label{eq.fit}
P(x) \sim \frac{1}{x^\alpha}, \quad \textrm{with} \quad
\alpha=2.41 \pm 0.12.
\end{equation}
This decay is in clear contrast to the exponential decay for the
monochromatic incident field. Note that within our range of
parameters a fit with $1/(x^\alpha \sqrt{\ln(x/\ell_0)})$ is not
noticeably different from the pure power law fit of Eq.
(\ref{eq.fit}). We have also checked that varying the atomic
density in the observation cell from $7\times 10^{16}$ to $2\times
10^{17}$~m$^{-3}$ has no impact on the measured value of $\alpha$.

\begin{figure}[t]
\begin{center}
\includegraphics{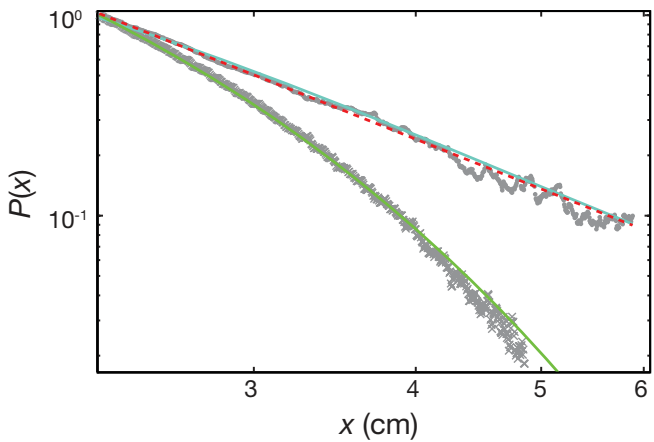}
\caption{{\bf Single step size distributions $P(x)$ plotted in
log-log scale.} For an incident monochromatic laser beam of
frequency $\omega$ (crosses), $P(x)$ shows an exponential
decrease, as shown by the green continuous fitting line. For an
incident Doppler-broadened field originated from a first
scattering cell (dots), $P(x)$ has a power law decrease, well
fitted by $P(x) \sim 1/x^\alpha$ with $\alpha=2.41 \pm 0.12$ (red
dashed line), characteristic of L\'evy flights. A microscopic
model, taking into account all hyperfine levels and Raman
transitions, is shown as the blue continuous line. Without any
free parameter except the vertical intensity scale, the agreement
with the experimental data is very satisfactory.}
\end{center}
\end{figure}

This result calls for two important comments. First, the exponent
$\alpha$ is smaller than $3$. We can thus not ignore the heavy
tails of the step size distribution. The variance of $P(x)$ is
infinite, making this distribution characteristic of L\'evy
flights. Second, we measure an important difference compared to
the prediction $\alpha=2$ expected when the natural width $\Gamma$
of the atomic transition is neglected [Eq. (\ref{eq.P2})]. Even
though this width $\Gamma/2\pi=6$~MHz is much smaller than the
Doppler broadening $\Delta\omega_\mathrm{D}/2\pi\sim 220$~MHz, one
cannot neglect the effect of the natural Lorentzian line shape on
the atomic absorption profile. In the opposite limit of negligible
Doppler broadening ($\Delta\omega_\mathrm{D}\ll \Gamma$) one
expects to recover the cold-atom limit, where all momenta of the
step size distribution are finite. It is thus not surprising that
the actual case of a finite natural width does increase the
absolute value of the exponent from the ideal Doppler limit, where
$\alpha=2$. A power law fit to the step size distribution $P(x)$
obtained by including this finite natural width in the numerical
integration of Eq. (\ref{eq:P}) yields an exponent of $2.3$, close
to the experimental value. We have also implemented a more refined
model, taking into account all hyperfine levels of rubidium atoms
including the hyperfine Raman transitions (see the Methods
section). The result of this numerical calculation is in excellent
agreement with the experimental data as shown in Fig. 3. Although
the finite natural line width is only a few percents of the
Doppler width, it surprisingly changes the power law exponent by
$15\%$. The measured difference to $\alpha=2$ has a particular
significance, as for $\alpha \leq 2$, even the mean free path is
no longer defined, in contrast to $\alpha > 2$ where $\langle x
\rangle$ is still finite.

The knowledge of the single step size distribution for photons
which have been scattered only once by hot atoms is however not
sufficient to describe the multiple scattering regime. Indeed, the
spectral characteristics of the photons depend on their previous
history. For instance, at angles close to forward scattering, the
Doppler broadening is very small and the scattering can be
considered to be almost elastic. This feature is described by a
partial frequency redistribution of the photons during their
multiple scattering process \cite{Molisch} and can be understood
as a memory effect in the random walk sequence.
In order to have
experimental access to the step size distribution (and therefore
to its exponent in case of power laws) in a multiple scattering
regime, it is important to extract the single step size
distribution of a photon that has been scattered several times by
hot atoms.

\begin{figure}[t]
\begin{center}
\includegraphics{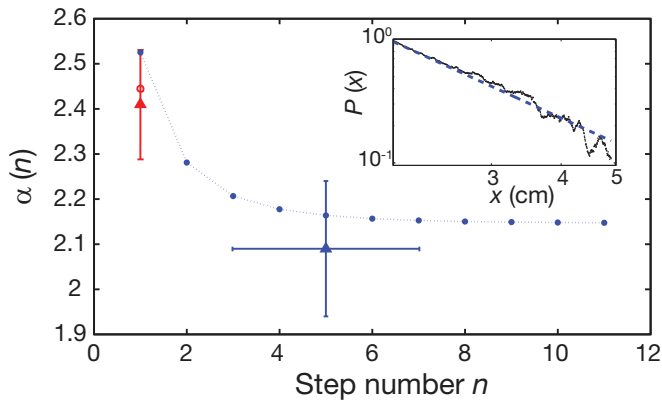}
\caption{{\bf Evolution of the power law exponent.} The double
cell configuration ($n=1$) [Fig. 1] yields at $90^\circ$ a pure
Doppler-broadened spectrum and the measured $\alpha$ (red
triangle) is $\alpha=2.41 \pm 0.12$. For the triple cell
configuration, photons are scattered about 5 times before entering
the observation cell. The measured exponent is then $\alpha=2.09
\pm 0.15$ (blue triangle). The data and fit for this last
configuration are shown in the inset. Vertical error bars
represent the uncertainty of the power law fit. The blue dots are
numerically computed with the real atomic structure and an angular
average of the Doppler-broadened frequency spectrum. For
comparison with the experiment, the open red circle is computed
without angular average.}\label{fig.alpha_n}
\end{center}
\end{figure}

To determine numerically the shape of the step size distribution
$P_n$ for photons after $n$ scattering events, we compute the
evolution of the spectrum $\Theta_n$ of scattered light, taking
into account Doppler broadening and averaging over the scattering
angles \cite{Portugais2}. $P_n$ is then obtained by using Eq.
(\ref{eq:P}) with $\Theta_n$ (see the Methods section). The
numerics show that $P_n$ quickly converges towards a power law,
which is independent of the initial frequency, and hence, that the
step size distribution of light in the multiple scattering regime
is well defined. The corresponding power law exponent $\alpha(n)$
is reported on Fig. \ref{fig.alpha_n}.

In order to approach experimentally the situation of multiple
scattering, we have implemented a triple cell geometry. In a first
cell, with an on-resonance optical thickness on the order of $2$,
we prepare photons that have been scattered several times ($n \sim
4$) when they leave the cell. Those photons have thus no memory of
their initial direction and their properties are well described by
an angular average of the fluorescence spectrum. These photons are
then sent towards the ``source cell" of Fig. 1, which still has a
low optical thickness. Photons thus undergo one more scattering
event with a well defined position before they can arrive onto the
observation cell. This experimental protocol allows us to produce
photons which have done multiple scattering and forgotten their
initial direction and frequency, as required for a steady state
situation. We then record the image of the fluorescence on the
CCD. The signal in this geometry is much weaker than in the
previous double cell geometry and is limited by photon shot noise
and cosmic rays. From the median averaging of 6 images
corresponding to an exposure time of 5 hours each, we are still
able to extract the step size distribution $P(x)$ after multiple
scattering and obtain a value $\alpha=2.09\pm 0.15$ [Fig. 4]. This
value, clearly below $3$, excludes a diffusion approach to be used
for multiple scattering of light in hot atomic vapours. This
experimental result is in good agreement with our numerical
estimation of the power law, taking into account all details of
the atomic transition [Fig. 4].

To conclude, we have directly measured the step size distribution
of photons undergoing a random walk in hot atomic vapours. Despite
a memory effect due to partial frequency redistribution, the step
size distribution of photons in hot atomic vapours converges after
a few steps to a power law with a diverging second moment.
Therefore photon trajectories in such a system are L\'evy flights.
The experiments described in this article show how it is possible
to obtain direct information on the microscopic ingredient leading
to superdiffusion of light in hot atomic vapours and open the path
to study the role of other broadening mechanisms, such as pressure
broadening or inelastic scattering at large intensities, where the
absorption cross sections still have Lorentzian wings, as for
natural broadening. However, in these cases, the emission spectra
are very different as one no longer has coherent emission
processes, and one expects in that case that $P(x)$ follows a
power law with an exponent $1.5$ leading to an infinite mean free
path \cite{Pereira}. Finally, truncated L\'evy flights
\cite{Stanley} could be considered to deal with finite-size
samples or, on the other hand, with very large systems, if the
step size distribution exhibits a cutoff at long distance
\cite{Portugais2}.

\ \\

\large

\noindent {\bf Methods}

\small

\noindent {\bf Correction of multiply scattered light.} The
required dynamic to determine the asymptotic behavior of $P(x)$
needs a fine trade-off between the amount of single scattering and
the need to avoid multiple scattering in the observation cell.
This is adjusted by the cell temperature and the subsequent atomic
density. At $41^\circ$C, the temperature used for our
measurements, this density is about $7\times 10^{16}$~m$^{-3}$, as
deduced from the laser beam attenuation. The mean free path for
resonant photons is then $\sim 9$~mm, and the effect of multiple
scattering cannot be totally neglected. It can however be
corrected. The intensity measured along the axis of the incident
light beam can indeed be written $I(x,0) = I_1(x,0) +I_{n \geq
2}(x,0)$, where $I_1$ is the intensity due to single scattering of
ballistic photons, $I_{n\geq 2}$ the one due to multiple
scattering. Slightly off the axis, however, only $I_{n\geq 2}$
remains. Assuming a smooth variation of $I_{n\geq 2}(x,d)$ at
distances $d$ much smaller than the mean free path of resonant
light we use $I_2(x,d)~\simeq~I_2(x,0)$ to subtract higher order
scattering along the center of the cell. Hence, we get
$P(x)\propto I_1(x,0) \propto I(x,0)-I(x,d)$. Step size
distribution measurements are thus obtained by subtracting
intensity signals observed on and off the ballistic beam axis. A
Monte Carlo simulation, which can track the position of a photon
emerging from the observation cell and register the number of
scattering it has performed, confirmed that this procedure
efficiently filters
single scattering events in the observation cell.\\

\noindent {\bf Numerical calculation of the step size distribution
for real atoms.} The step size distribution is computed from Eq.
(\ref{eq:P}) with the knowledge of the frequency-dependent
mean-free path $\ell(\omega)$ (the inverse of which is the
absorption spectrum) and the emission spectrum $\Theta(\omega)$.
The mean free path at frequency $\omega$ is given by
\begin{equation}
\label{eq:l} \frac{1}{\ell(\omega)} =
n_{at}\int_{-\infty}^{+\infty} \sigma
\left(\omega(1-\frac{v_x}{c})\right) P_\mathrm{M,1}(v_x) dv_x \; ,
\end{equation}
where $n_{at}$ is the atomic density, $P_\mathrm{M,1}$ is the
Maxwell distribution of atomic velocities $v_x$ along the
direction of light propagation $x$ in the observation cell, and
$\sigma$ is the atomic cross-section, taken at the Doppler-shifted
frequency $\omega(1-\frac{v_x}{c})$. The emission spectrum in the
first cell is given by
\begin{eqnarray}\label{eq:spectra1}
\Theta_1(\omega) \propto && \int_{0}^{+\infty} d\omega'
\int_{-\infty}^{+\infty}dv_x \int_{-\infty}^{+\infty}dv_y \,
\Theta_0(\omega')\, \sigma \left( \omega'(1-\frac{v_y}{c}) \right)
\nonumber \\ && \times~P_\mathrm{M,2}(v_x,v_y) \,
\delta\left(\omega -
\omega'(1-\frac{v_y}{c})(1+\frac{v_x}{c})\right) \; ,
\end{eqnarray}
where $\Theta_0(\omega')$ is the spectrum of the incident laser
propagating along $y$,
$\omega_\mathrm{at}=\omega'(1-\frac{v_y}{c})$ is the incident
photon frequency in the atomic rest frame for atoms at velocity
$v_y$ along $y$, and $\omega_\mathrm{at}(1+\frac{v_x}{c})$ is the
frequency in the laboratory frame of the photon emitted along $x$,
towards the second cell. $P_\mathrm{M,2}$ is the Maxwell
distribution of atomic velocities along two directions. The Dirac
distribution inside the integral denotes the energy conservation
during \emph{coherent} scattering, the only change in frequency
coming from Doppler shifts. We take as the laser spectrum
$\Theta_0$ a Lorentzian of width $0.6$~MHz. With two-level atoms,
the cross-section $\sigma$ is Lorentzian-shaped (natural width
$\Gamma/2\pi=6.07$~MHz) and we obtain the first order step size
distribution by reporting Eqs.
(\ref{eq:Pxw},\ref{eq:l},\ref{eq:spectra1}) into Eq. (\ref{eq:P}).
Taking into account all the levels of rubidium atoms is done by
writing the scattering cross-section as the sum of all the
possible transitions weighted by their respective strength factors
and with the hypothesis of an equipartition of atomic population
among all possible states.

The multiple scattering regime is characterized by the
$n^\mathrm{th}$ order step size distribution function, which is
computed by using $\Theta_n(\omega)$ instead of $\Theta(\omega)$
in Eq. (\ref{eq:P}). The evolution of the spectrum, taking into
account Doppler broadening and averaging over the scattering
angles \cite{Portugais2}, is then given by
\begin{equation}\label{eq:spectreEvolution}
\Theta_{n}(\omega) = \int_{-\infty}^{+\infty}
\Theta_{n-1}(\omega') \frac{R(\omega,\omega')}{\Phi(\omega')} \,
d\omega' \; ,
\end{equation}
where $\Phi(\omega') \propto 1/\ell(\omega')$ is the absorption
spectrum and $R$ is the joint laboratory frame redistribution
function of the scatterers, which gives the probability for a
photon at a frequency $\omega'$ to be scattered at a frequency
$\omega$ \cite{Molisch}. We take for the redistribution function
$R$ the average over all the possible transitions of the standard
redistribution function of a Voigt absorption profile
\cite{Molisch}.

\large  \noindent {\bf Acknowledgements}

\small \noindent We acknowledge financial support from the program
ANR-06-BLAN-0096 and funding for N.M. by DGA.


\begin{thebibliography}{99}

\bibitem{Levy} L\'evy, P. {\it Theorie de l'Addition des Variables Aleatoires} (Gauthier-Villiers, Paris, 1937).
\bibitem{Bouchaud} Bouchaud, J.-P. and Georges, A. Anomalous diffusion in disordered media: Statistical mechanisms, models and physical applications. {\it Phys. Rep.} {\bf 195}, 127--293 (1990).
\bibitem{Shlesinger} Shlesinger, M., Zaslavsky, G., Frisch, U., Eds., {\it L\'evy Flights and Related Topics in Physics} (Springer-Verlag, New York, 1995).
\bibitem{Metzler} Metzler, R. and Klafter, J. The random walk's guide to anomalous diffusion: a fractional dynamics approach. {\it Phys. Rep.} {\bf 339}, 1--77 (2000).
\bibitem{Bouchaud2} Bouchaud, J.-P. and Potters, M. {\it Theory of Financial Risk and Derivative Pricing} (Cambridge University Press, 2003).
\bibitem{Botet} Botet, R. and Ploszajczak, M. {\it Universal Fluctuations} (World Scientific, Singapore, 2002).
\bibitem{Goldenfeld} Goldenfeld, N. D. {\it Lectures on Phase Transitions and the Renormalisation Group} (Addison-Wesley, 1992).
\bibitem{Wiersma} Barthelemy, P., Bertolotti, J., and Wiersma, D. S. A L\'evy flight for light. {\it Nature} {\bf 453}, 495--498 (2008).
\bibitem{Springmann} Springmann, U., Multiple resonance line scattering and the ``momentum problem'' in Wolf-Rayet star winds. {\it Astron. Astrophys.} {\bf 289}, 505--523 (1994).
\bibitem{Molisch} Molisch, A. F. and Oehry, B. P., {\it Radiation Trapping in Atomic Vapours} (Oxford University, Oxford, 1998).
\bibitem{Holstein} Holstein, T. Imprisonment of Resonance Radiation in Gases. {\it Phys. Rev.} {\bf 72}, 1212--1233 (1947).
\bibitem{Kenty} Kenty, C. On Radiation Diffusion and the Rapidity of Escape of Resonance Radiation from a Gas. {\it Phys. Rev. } {\bf 42} 823--842 (1932) .
\bibitem{Fioretti} Fioretti, A., Molisch, A. F., Mutter, J. H., Verkerk, P. and Allegrini, M. Observation of radiation trapping in a dense Cs magneto-optical trap. {\it Opt. Commun.} {\bf 149}, 415--422 (1998).
\bibitem{Labeyrie2003} Labeyrie, G. {\it et al.} Slow Diffusion of Light in a Cold Atomic Cloud. {\it Phys. Rev. Lett.} {\bf 91}, 223904 (2003).
\bibitem{Labeyrie2005} Labeyrie, G., Kaiser, R. and Delande, D. Radiation trapping in a cold atomic gas. {\it Appl. Phys. B} {\bf 81}, 1001--1008  (2005).
\bibitem{Pereira} Pereira, E. , Martinho, J. M. G. and Berberan-Santos, M. N. Photon Trajectories in Incoherent Atomic Radiation Trapping as L\'evy Flights. {\it Phys. Rev. Lett.} {\bf 93}, 120201 (2004).
\bibitem{Portugais2} Alves-Pereira, A. R., Nunes-Pereira, E. J., Martinho, J. M. G. and Berberan-Santos, M. N. Photonic superdiffusive motion in resonance line radiation trapping Partial frequency redistribution effects. {\it J. Chem. Phys.} {\bf 126}, 154505 (2007).
\bibitem{Stanley} Mantegna, R. N. and Stanley, H. E. Stochastic Process with Ultraslow Convergence to a Gaussian: The Truncated L\'evy Flight. {\it Phys. Rev. Lett.} {\bf 73}, 2946--2949 (1994).
\end{thebibliography}
\end{document}